\documentclass{article}
\usepackage{spconf,amsmath,graphicx}

\usepackage{enumitem}
\setlist{nosep, leftmargin=14pt}

\usepackage{mwe} 

\title{Identifying virulence determinants in pathogenic mycobacteria via changes in host cell mitochondrial morphology}
\name{Shannon Quinn$^{1,2,*,\dagger}$, Amr Abbadi$^{3,*}$, Seyed Alireza Vaezi$^{1}$, Russell K. Karls$^{3}$, Frederick D. Quinn$^{3,\dagger}$\thanks{$^{*}$ Indicates equal contributions \newline $^{\dagger}$ Indicates corresponding authors}}
\address{University of Georgia \\
	$^{1}$ School of Computing \\
	$^{2}$ Department of Cellular Biology \\
	$^{3}$ Department of Infectious Diseases \\
	Athens, GA, United States}
\begin{document}
%
\maketitle
\begin{abstract}

The goal of this study is to develop a computational model of the progression of changes in mitochondrial phenotype resulting from infection with pathogenic mycobacteria. This ultimately will enable a large-scale virulence screen of mutant bacterial libraries. \emph{Mycobacterium tuberculosis} (\emph{Mtb}) is an intracellular pathogen, but only a small number of its genes have been studied for roles in intracellular host cell survival and replication. Mitochondria are the powerhouse of the host cell and play critical roles in cell survival when attacked by certain pathogens. When \emph{Mtb} bacteria invade host cells, they induce changes in mitochondrial morphology, making mitochondria a novel target for image processing and machine learning to determine virulence associations of genes in \emph{Mtb} and potentially other related intracellular pathogens. By hypothesizing mitochondria as an instance of a dynamic and interconnected graph, we demonstrate a statistical approach for quantitatively recognizing novel mitochondrial phenotypes induced by invading pathogens.
\end{abstract}
\begin{figure}
	\centering
	\includegraphics[height=2in]{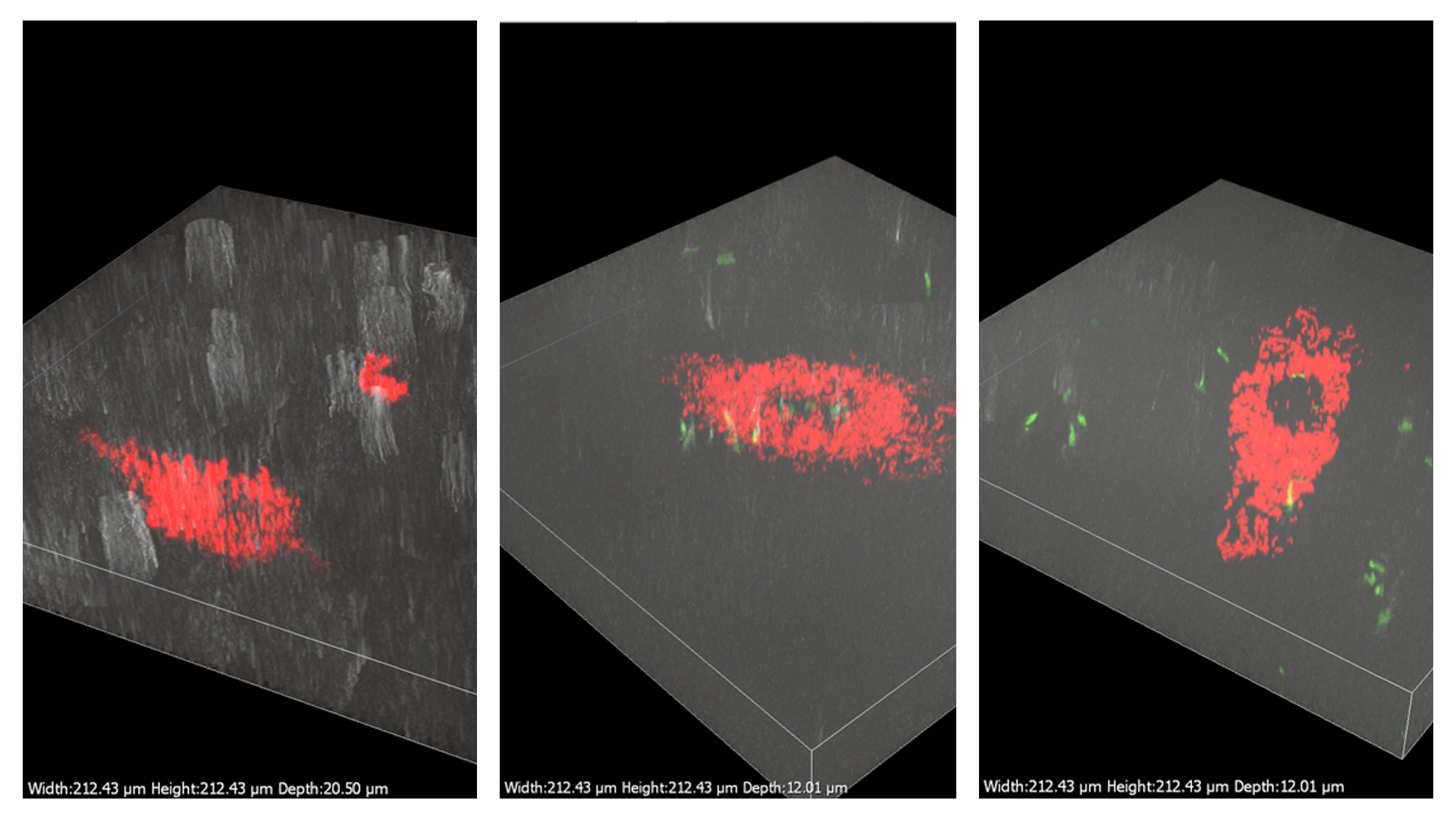}
	\caption{Composite Z-stack images of A549 cells producing DsRed mitochondria at a single time point. Red indicates mitochondria, green indicates \emph{Mmar} bacteria. Cells were uninfected (\textbf{left} panel) or infected with GFP-expressing wild type \emph{Mmar} (\textbf{center}) or ESAT-6 operon mutant (\textbf{right}).}
	\label{fig:data}
\end{figure}

\vspace{-1em}
\section{Introduction}
\label{sec:intro}

\emph{Mtb} kills over 1 million people each year~\cite{mai2024exposure}. During infection of particular cell types, \emph{Mtb} induces host cell death programs; however, the specific mycobacterial mechanisms involved are only partially understood. The ultimate goal is to identify virulence genes responsible for changes in host cell morphology of invaded cells over time by screening mycobacterial transposon mutant libraries. This would entail not only high-resolution, real-time fluorescence microscopy in 2D and 3D, but also also computational models that capture changes in organellar morphology as a function of gene knockout status and infection stage.

This project is motivated by an observation of mitochondrial behavior in alveolar epithelial cells following invasion by \emph{Mtb} bacteria~\cite{fine2015infection}: mitochondria are typically spread diffusely across the interior of the cell and migrate away from the invading pathogen before the cell is killed. This observation postulated a number of research questions, such as: what molecular machinery within \emph{Mtb} is responsible for inducing this phenomenon? Do other pathogens induce this response? Are there quantifiable patterns for the host's response?

\begin{figure}
	\centering
	\includegraphics[height=3.5in]{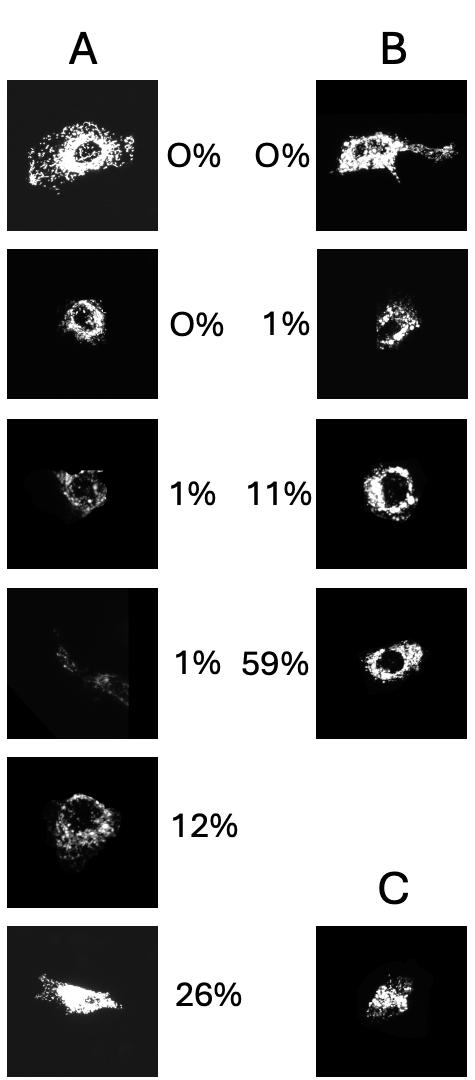}
	\caption{The 11 cells in our data most consistently misclassified, and respective rates at which our model \emph{correctly} classified them. \textbf{A}: Mutant infected cells, and \textbf{B}: wildtype infected cells. \textbf{C}: The single uninfected control that is classified as wildtype, rather than the expected mutant.}
	\label{fig:exemplars}
\end{figure}

In this phase of the project, we are building the computational tools for measuring mitochondrial behavior and generating data-driven hypotheses. Data generation entails culturing mutant and wild type strains of \emph{Mycobacterium marinum} (\emph{Mmar}), a less virulent close relative of \emph{Mtb}, that can infect and kill the same host cells. The mitochondria and \emph{Mmar} bacteria are fluorescently tagged for downstream analyses. The computational steps involve image processing and modeling of 3D fluorescent confocal time series data. Previous studies have shown that knocking out the gene in \emph{Mtb} encoding protein ESAT-6 results in normal mitochondrial appearance and subcellular spatial distributions (morphology), providing strong evidence that ESAT-6 plays an important role in inducing morphological changes in mitochondria upon invasion by \emph{Mtb}~\cite{fine2015infection}. The ESX1 secretion system, of which ESAT-6 is an essential component,  is highly conserved and mechanistically similar in \emph{Mmar}~\cite{chirakos2020modeling}, suggesting the same mitochondrial behavior is likely in \emph{Mmar}-infected cells. As proof-of-principle, mitochondrial morphology is examined in host cells infected with wildtype (wt) \emph{Mmar} versus an ESAT-6 operon mutant (mu).

\vspace{-1.5em}
\section{Related Work}
\label{sec:related}

The development of automated segmentation, tracking, and analysis of diffuse or punctuate subcellular structures has been uneven. Unlike identifying nuclei or even cells, structures largely presenting with closed and convex shapes, mitochondria are an organelle of varying size, shape, and spatial distribution. In previous work, we hypothesized the mitochondrial phenotype as an instance of a ``social network,'' consisting of connected and interacting nodes~\cite{fazli2020ornet}. Changes in the phenotype corresponded to the forming and breaking of connections between nodes~\cite{hill2020spectral}. This representation was effective in recognizing changes in mitochondrial morphology due to fission or fusion events~\cite{pulagam2021classification}, but the connection between the high-level graphical model and the underlying biology was tenuous.

Others have used the same graph-theoretic modeling hypothesis to capture quantitative metrics~\cite{chu2022image}. Mitometer~\cite{lefebvre2021automated} and Mito Hacker~\cite{rohani2020mito} offer similar functionality of segmenting and tracking individual mitochondrial nodes and deriving quantitative metrics, but both tools have been unmaintained for some time. MitoTNT ~\cite{wang2023mitotnt} provides extensive temporal tracking capabilities for mitochondrial dynamics, and it uses MitoGraph~\cite{harwig2018methods} to include automated segmentation and quantitative analysis of mitochondrial morphology.

\vspace{-1.75em}
\section{Methods}
\label{sec:methods}

\begin{figure*}
	\centering
	\includegraphics[height=1.93in]{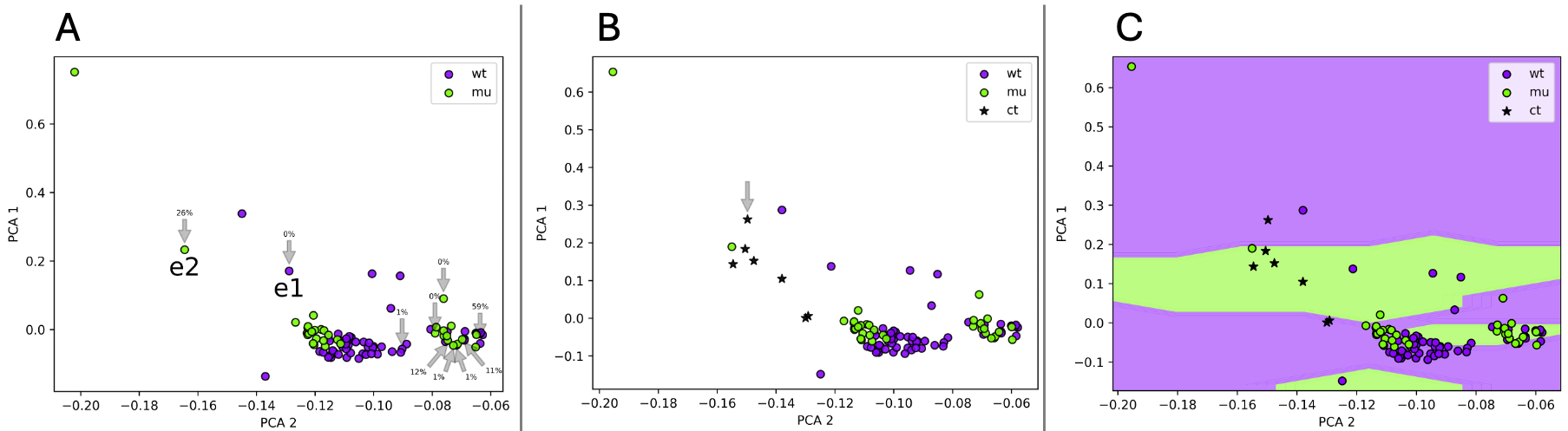}
	\caption{Our data embedded in 2D PCA space, where each dot is a single cell. \textbf{A}: The top 10 most misclassified cells (4 wt, 6 mu) are highlighted by arrows, with percentages indicating how often they are correctly classified. \texttt{e1} is wt but is classified as such 0\% of the time, whereas \texttt{e2} is mu and classified as such 26\% of the time. \textbf{B}: The ct are shown, with an arrow indicating the one that is classified as wt (all others are classified as mu). \textbf{C}: The classifier's decision space is underlaid the dots, highlighting where misclassifications happen.}
	\label{fig:pca}
\end{figure*}

\vspace{-1em}
\subsection{Data}

Wildtype \emph{Mmar} strain M and ESAT-6 operon mutant Tn8064 (MMAR\_5451 insertion mutant) transformed with plasmid pRK295 (pMV261~\cite{stover1991new} encoding green-fluorescent protein (GFP) variant GFPmut2~\cite{cormack1996facs}) were maintained in standard culture medium (Middlebrook 7H9tgADS~\cite{braunstein2002genetic} with kanamycin (0.02 mg/mL). Human alveolar epithelial cell line A549 transformed with plasmid pDsRed2-Mito-7 were periodically treated with geneticin (1 mg/mL) to maintain cells producing DsRed2-labeled mitochondria (A549-DsRed). A549-DsRed cells cultured in DMEM medium supplemented with 10\% fetal bovine serum were seeded into 96-well glass-bottomed dishes and infected with \emph{Mmar} strains at a multiplicity of infection of 100 for 24 hours, treated with amikacin (0.2 mg/mL) for 2 hours, washed 3X with culture medium, and incubated at 35$^{o}$C in an environmental chamber on a Nikon Eclipse Ti confocal microscope for time-lapse image capture at 30 minute intervals for 16 hours at 60X magnification. Lasers and detectors were set to capture fluorescence at 488m (red), 562.2m (green) and transmitted light to outline the host cells. Z-stacks were acquired at 0.5 $\mu$m intervals using a step motor to obtain 3D datasets (Fig.~\ref{fig:data}).

\vspace{-1em}
\subsection{Preprocessing}

We chose MitoTNT and MitoGraph for preprocessing and feature extraction. The z-stacks were annotated to crop and segment each cell. MitoGraph was used for mitochondrial segmentation and feature measurements. The extracted features include volume by voxels, average width, total length, and volume calculated from length~\cite{harwig2018methods}. As these features were calculated per-frame, we also computed the average, variance, median, minimum, and maximum values of these quantities per cell over the entire video. These features were aggregated into 30-length feature vectors, one per cell. In total, we computed feature sets for 99 cells (57 wt, 35 mu, and 7 ct). Put another way, the 3D z-stack time series confocal data was transformed to $X^{99 \times 30}$.

\vspace{-1em}
\section{Results}
\label{sec:results}

We trained a variety of binary classifiers to differentiate mutant from wildtype ($n = 92$), including linear models, decision trees, and nonlinear support vector machines. The models were trained with 5-fold stratified cross-validation (80\% of the data for training, 20\% for testing), repeated 100 times with different randomization. Our best model, the nonlinear support vector machine (NuSVC), reported an average \textbf{87.258\% accuracy ($\pm$ 6.50\%)}, besting the next model by nearly 10 full percentage points.

Over these 100 randomized cross-validation iterations, we examined the data that the NuSVC consistently struggled to correctly identify. We found 10 exemplars whose average correct classification rate was less than the average rate of the classifier itself. Fig.~\ref{fig:exemplars} lists these cells and the respective rates at which the model correctly classified them.

We use principal components analysis (PCA) to embed our data $X$ into a 2D space and further examine the NuSVC decision process. We found the top 2 PCA dimensions contained \textbf{96.788\%} of the total variance in the data, making it an excellent embedding space for further investigation. Fig.~\ref{fig:pca}A shows the embedded data, with arrows highlighting the 10 exemplars from Fig.~\ref{fig:exemplars} and their respective accuracy rates.

We also ran the 7 held-out ct cells through the trained NuSVC, and it recognized 6 of the 7 as mutant type; the 1 it recognized as wildtype is indicated in Fig.~\ref{fig:exemplars}C and Fig.~\ref{fig:pca}B. Finally, we evaluated the NuSVC decision process across the PCA space to better indicate where its internal boundaries between mu and wt were (Fig.~\ref{fig:pca}C).

\vspace{-1em}
\section{Discussion}
\label{sec:discussion}

First and foremost, the 87\%+ binary classification rate provides strong evidence of detectable changes in mitochondrial phenotype between infections from wt versus mu \emph{Mmar}. The graph-theoretic features computed from mitochondrial appearance performed very well in a situation where changes in phenotype were expected. Second, however, the extremely high amount of variance captured in only two PCA dimensions (96\%+) suggests subtle but strong underlying correlation between features, likely from the feature aggregation functions. A more diverse feature set would be prudent in larger-scale transposon mutant screens. Third, some combination of feature insensitivity, biological variance, or poor pathogen invasion is likely responsible for the mistakes of the classifier on specific cells. Cells in Fig.~\ref{fig:exemplars}A look like wt cells (due to feature insensitivity or biological variations), and likewise Fig.~\ref{fig:exemplars}B look like mu cells (due to poor infection rates by \emph{Mmar}, resulting in subtle or nonexistent changes to mitochondria). We would expect Fig.~\ref{fig:exemplars}C to look uninfected by virtue of the mutant \emph{Mmar} lacking a critical virulence gene, but it instead looks like wt. We also see this in Fig.~\ref{fig:pca}B (the arrow) and especially Fig.~\ref{fig:pca}C where that same ct point lies within the wt space of the classifier. The classifier's tendency to ``confuse'' uninfected cells as mutants provides additional quantitative evidence for the ESAT-6 gene's role as an \emph{Mtb} virulence trait~\cite{fine2015infection}.

\vspace{-0.2em}

A deep learning approach, while all but certain to attain a higher accuracy, would be much more difficult to interpret. For future work screening thousands of mutants whose effects on the organellar morphologies are unknown, it is critical to be able to explain significant morphological changes. Provided the model could maintain a certain degree of insensitivity to natural biological variations--no small feat, given deep models are notoriously overparameterized--this could be a possible way forward.

\vspace{-1em}
\section{Conclusion}
\label{sec:conclusion}

In this study, we have demonstrated that graph-theoretic features computed from fluorescently-tagged mitochondria in 3D time series confocal data can be very useful for identifying significant changes in mitochondrial phenotype. We achieved over 87\% accuracy in differentiating mitochondrial phenotypes of cells infected by wildtype \emph{Mmar} versus those infected by an ESAT-6 knockout mutant \emph{Mmar} variant. We also highlighted weaknesses in both the feature set and the classifier by visualizing low-dimensional embeddings of both. These findings provide a foundation for future large-scale mutant screens to further study virulence factors of \emph{Mmar} and \emph{Mtb}.

%


\vspace{-1.3em}
\section{Acknowledgments}
\label{sec:acknowledgments}

We thank Christina Cosma, Lalita Ramakrishnan (University of Cambridge, UK), and David Tobin (Duke University) for \emph{Mmar} strain Tn8064 (Tn5370-mediated insertion into MMAR\_5451). DsRed2-Mito-7 was a gift from Michael Davidson (Addgene plasmid \# 55838; http://n2t.net/addgene:55838 ; RRID:Addgene\_55838).

This work was supported by NIH grant 5R21AI151453-02. The authors declare no personal or financial conflicts of interest. This study relied on cell lines and numerical analyses, so no ethical approvals were required.

\vspace{-1.3em}
\bibliographystyle{IEEEbib}
\bibliography{Template_ISBI_latex.bib}

\begin{thebibliography}{10}

\bibitem{mai2024exposure}
Dat Mai, Ana Jahn, Tara Murray, Michael Morikubo, Pamelia~N Lim, Maritza~M Cervantes, Linh~K Pham, Johannes Nemeth, Kevin Urdahl, Alan~H Diercks, et~al.,
\newblock ``Exposure to mycobacterium remodels alveolar macrophages and the early innate response to mycobacterium tuberculosis infection,''
\newblock {\em PLoS Pathogens}, vol. 20, no. 1, pp. e1011871, 2024.

\bibitem{fine2015infection}
Kari Fine-Coulson, Steeve Gigu{\`e}re, Frederick~D Quinn, and Barbara~J Reaves,
\newblock ``Infection of a549 human type ii epithelial cells with mycobacterium tuberculosis induces changes in mitochondrial morphology, distribution and mass that are dependent on the early secreted antigen, esat-6,''
\newblock {\em Microbes and infection}, vol. 17, no. 10, pp. 689--697, 2015.

\bibitem{chirakos2020modeling}
Alexandra~E Chirakos, Ariane Balaram, William Conrad, and Patricia~A Champion,
\newblock ``Modeling tubercular esx-1 secretion using mycobacterium marinum,''
\newblock {\em Microbiology and Molecular Biology Reviews}, vol. 84, no. 4, pp. 10--1128, 2020.

\bibitem{fazli2020ornet}
Mojtaba Fazli, Marcus Hill, Andrew Durden, Rachel Mattson, Allyson~T Loy, Barbara Reaves, Abigail Courtney, Frederick~D Quinn, Chakra Chennubhotla, and Shannon Quinn,
\newblock ``Ornet-a python toolkit to model the diffuse structure of organelles as social networks,''
\newblock {\em Journal of Open Source Software}, vol. 5, no. 47, pp. 1983, 2020.

\bibitem{hill2020spectral}
Marcus Hill, Mojtaba~Sedigh Fazli, Rachel Mattson, Meekail Zain, Andrew Durden, Allyson~T Loy, Barbara Reaves, Abigail Courtney, Frederick~D Quinn, S~Chakra Chennubhotla, et~al.,
\newblock ``Spectral analysis of mitochondrial dynamics: A graph-theoretic approach to understanding subcellular pathology.,''
\newblock in {\em SciPy}, 2020, pp. 91--97.

\bibitem{pulagam2021classification}
Neelima Pulagam, Marcus Hill, Mojtaba~Sedigh Fazli, Rachel Mattson, Meekail Zain, Andrew Durden, Frederick~D Quinn, S~Chakra Chennubhotla, and Shannon~P Quinn,
\newblock ``Classification of diffuse subcellular morphologies.,''
\newblock in {\em SciPy}, 2021, pp. 115--122.

\bibitem{chu2022image}
Ching-Hsiang Chu, Wen-Wei Tseng, Chan-Min Hsu, and An-Chi Wei,
\newblock ``Image analysis of the mitochondrial network morphology with applications in cancer research,''
\newblock {\em Frontiers in Physics}, vol. 10, pp. 855775, 2022.

\bibitem{lefebvre2021automated}
Austin~EYT Lefebvre, Dennis Ma, Kai Kessenbrock, Devon~A Lawson, and Michelle~A Digman,
\newblock ``Automated segmentation and tracking of mitochondria in live-cell time-lapse images,''
\newblock {\em Nature Methods}, vol. 18, no. 9, pp. 1091--1102, 2021.

\bibitem{rohani2020mito}
Ali Rohani, Jennifer~A Kashatus, Dane~T Sessions, Salma Sharmin, and David~F Kashatus,
\newblock ``Mito hacker: a set of tools to enable high-throughput analysis of mitochondrial network morphology,''
\newblock {\em Scientific reports}, vol. 10, no. 1, pp. 18941, 2020.

\bibitem{wang2023mitotnt}
Zichen Wang, Parth Natekar, Challana Tea, Sharon Tamir, Hiroyuki Hakozaki, and Johannes Sch{\"o}neberg,
\newblock ``Mitotnt: Mitochondrial temporal network tracking for 4d live-cell fluorescence microscopy data,''
\newblock {\em PLoS computational biology}, vol. 19, no. 4, pp. e1011060, 2023.

\bibitem{harwig2018methods}
Megan~C Harwig, Matheus~P Viana, John~M Egner, Jason~J Harwig, Michael~E Widlansky, Susanne~M Rafelski, and R~Blake Hill,
\newblock ``Methods for imaging mammalian mitochondrial morphology: A prospective on mitograph,''
\newblock {\em Analytical biochemistry}, vol. 552, pp. 81--99, 2018.

\bibitem{stover1991new}
CK~Stover, VF~De~La~Cruz, TR~Fuerst, JE~Burlein, LA~Benson, LT~Bennett, GP~Bansal, JF~Young, MH~Lee, GF~Hatfull, et~al.,
\newblock ``New use of bcg for recombinant vaccines,''
\newblock {\em Nature}, vol. 351, no. 6326, pp. 456--460, 1991.

\bibitem{cormack1996facs}
Brendan~P Cormack, Raphael~H Valdivia, and Stanley Falkow,
\newblock ``Facs-optimized mutants of the green fluorescent protein (gfp),''
\newblock {\em Gene}, vol. 173, no. 1, pp. 33--38, 1996.

\bibitem{braunstein2002genetic}
Miriam Braunstein, Stoyan~S Bardarov, and William~R Jacobs~Jr,
\newblock ``Genetic methods for deciphering virulence determinants of mycobacterium tuberculosis,''
\newblock in {\em Methods in enzymology}, vol. 358, pp. 67--99. Elsevier, 2002.

\end{thebibliography}

\end{document}